\documentclass[sigconf]{acmart}

\usepackage{graphicx}
\usepackage{enumitem}
\usepackage{tabularx}
\usepackage{makecell}
\usepackage{booktabs}
\usepackage{multirow}
\usepackage{xcolor}
\usepackage{caption}
\usepackage{subcaption}
\usepackage{setspace}
\usepackage{wrapfig}
\graphicspath{{img/}}

\newcommand{\paragraphb}[1]{\vspace{0.03in}\noindent{\bf #1} }
\newcommand{\subscript}[2]{$#1 _ #2$}

\title{WaveVerif: Acoustic Side-Channel based Verification of Robotic Workflows}

\author{Zeynep Yasemin Erdogan\\\footnotesize (Corresponding author)}
\affiliation{%
  \institution{Newcastle University}
  \country{United Kingdom}}
\email{z.y.erdogan2@newcastle.ac.uk}

\author{Shishir Nagaraja}
\affiliation{%
  \institution{PES University}
  \country{India}}
\email{shishir.nagaraja@pes.edu}

\author{Chuadhry Mujeeb Ahmed }
\affiliation{%
  \institution{Newcastle University}
  \country{United Kingdom}}
\email{mujeeb.ahmed@newcastle.ac.uk}

\author{Ryan Shah}
\affiliation{%
  \institution{Sapphire}
  \country{United Kingdom}}
\email{ryan.k.shah@gmail.com}

\begin{document}

\begin{abstract}
In this paper, we present a framework that uses acoustic side-channel analysis (ASCA) to monitor and verify whether a robot correctly executes its intended commands. We develop and evaluate a machine-learning-based workflow verification system that uses acoustic emissions generated by robotic movements. The system can determine whether real-time behavior is consistent with expected commands. The evaluation takes into account movement speed, direction, and microphone distance. The results show that individual robot movements can be validated with over 80\% accuracy under baseline conditions using four different classifiers: Support Vector Machine (SVM), Deep Neural Network (DNN), Recurrent Neural Network (RNN), and Convolutional Neural Network (CNN). Additionally, workflows such as pick-and-place and packing could be verified with similarly high confidence. Our findings demonstrate that acoustic signals can support real-time, low-cost, passive verification in sensitive robotic environments without requiring hardware modifications.

\end{abstract}


\maketitle


\section{Introduction}
\label{sec:introduction}

The proliferation of 
robotic systems across diverse domains—including industrial automation, healthcare, logistics, education, agriculture, and telepresence—has led to substantial improvements in productivity, precision, and operational efficiency \cite{aschenbrenner2015teleoperation, bartos2021automotive, murphy2000}. As these systems become increasingly interconnected and embedded within broader cyber-physical infrastructures, ensuring their security and operational integrity has emerged as a critical challenge. While extensive research has focused on mitigating threats posed by active adversaries—such as command injection, software manipulation, and data channel tampering 
\cite{ahmed2020challenges,humphreys2008}—comparatively less attention has been devoted to post-command verification mechanisms that assess the correctness of a robot’s physical behavior following command execution~\cite{ahmed2024time}.

In this work, we introduce a passive, non-invasive verification framework that leverages Acoustic Side-Channel Analysis (ASCA)~\cite{sokasca} to assess whether a robot physically executes its assigned tasks as intended. The sounds generated by actuators, motors, and mechanical interactions encapsulate rich temporal and spectral features that reflect the robot’s motion dynamics and operational patterns \cite{shah2022fingerprinting}. This approach is particularly valuable in scenarios where internal telemetry data may be unreliable due to adversarial threats, sensor spoofing, or communication disruptions. In such contexts, a defender can deploy an external recording device—such as a smartphone or dedicated microphone—near the robot to capture its acoustic emissions. These recordings are then compared with pre-trained models that represent legitimate behavioural signatures. By matching the observed audio to known acoustic fingerprints associated with specific commands, the system can detect deviations that may indicate manipulation, malfunction, or unauthorised alterations in behaviour.

To evaluate the robustness of our approach, we collect acoustic emissions across a range of robotic tasks, including fundamental 3D-axis movements and more complex workflows such as pick-and-place and packaging operations. Using this dataset, we train inference models to classify motion patterns under varying conditions, including different speeds, distances, and microphone placements. We evaluate both deep learning and non-neural models to compare their performance across the considered experimental conditions.


Our results demonstrate that acoustic emissions are a reliable and effective tool to verify robotic behaviour. Under baseline conditions, our system succeeds with 80\% or more accuracy (across four classifiers) in classifying individual movement patterns and maintains strong performance even when parameters are varied. These findings highlight the potential of acoustic side channels, traditionally viewed as a vulnerability, to provide a practical basis for external, real-time verification of robotic actions. This approach provides a lightweight and hardware-independent verification mechanism that improves the reliability of critical robotic systems, especially when operational accuracy and transparency are important.

The remainder of this paper is organised as follows: Section~\ref{sec:background} provides background on teleoperated robotics and acoustic emanations, followed by a description of the threat model. Section~\ref{sec:defence_evaluation} outlines our proposed verification methodology and presents the experimental results. In Section~\ref{sec:discussion}, we offer an in-depth discussion of the implications and limitations of our approach. Related work is reviewed in Section~\ref{sec:related}, and we conclude in Section~\ref{sec:conclusion}.
\section{Motivation and Goals}
\label{sec:background}

\begin{figure}[t]
	\centering
	\includegraphics[width=0.6\linewidth]{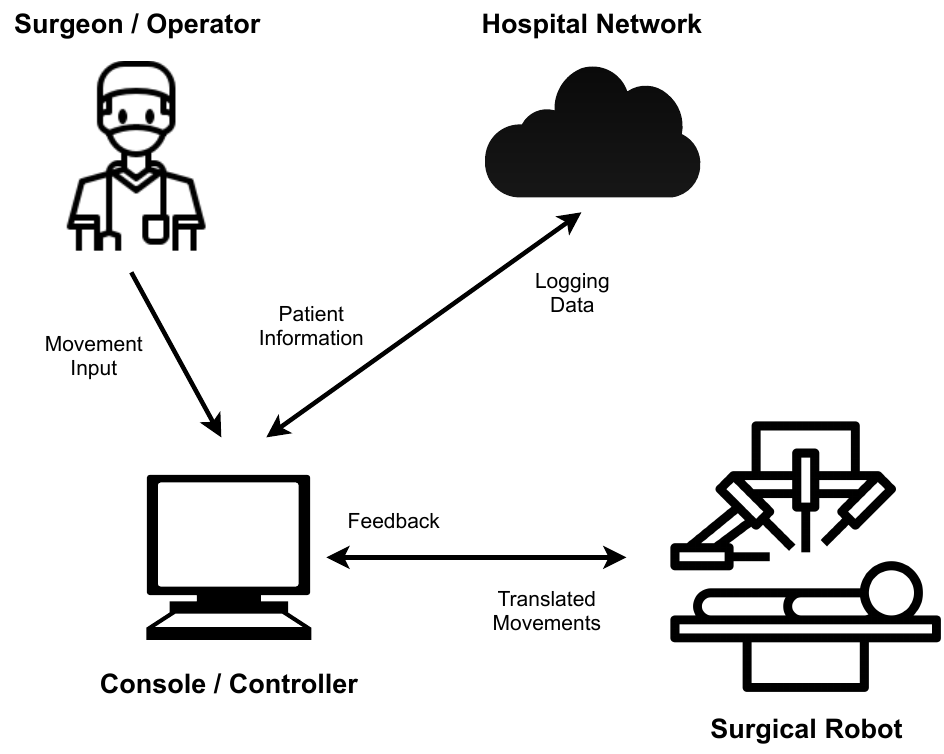}
	\caption{Teleoperated Robot Architecture.}
 \vspace{-0.5 cm}
	\label{fig:teleoparch}
\end{figure}

\subsection{Networked Robots}
Connected robots are extensively deployed across a range of sectors, including manufacturing, logistics, defense, and healthcare. These systems enable human operators to control robotic agents (often remotely) and have become indispensable in applications such as structural health monitoring, warehouse automation and telesurgery \cite{aschenbrenner2015teleoperation,avila2020teleoperated,bartos2021automotive,grabowski2021teleoperated,hannaford2012raven,li2017teleoperation,sung2001laparoscopic,tewari2002prostatectomy}. A typical connected robot architecture comprises two primary subsystems: a master interface, where the operator issues commands, and a slave robotic unit that executes the corresponding physical actions. These components are interconnected via an electronic control system and often communicate over organizational networks or external communication links. A typical robot architecture is illustrated in Figure~\ref{fig:teleoparch}.

In surgical applications, this architecture allows expert surgeons to perform procedures, with the robotic system executing commands in real time \cite{marescaux2006telesurgery,okamura2004haptic}. In industrial settings, operated robots carry out repetitive or hazardous tasks with high precision. However, the reliability of these systems is paramount, as any discrepancy between the intended and actual movements can compromise safety, privacy, or operational objectives. Ensuring the integrity of robotic behavior is therefore critical. While traditional cybersecurity measures focus on securing communication channels and controller logic, they often lack visibility into the robot’s physical execution. This limitation motivates the exploration of external, passive verification techniques—such as those based on acoustic emissions—to enhance trust in robotic operations.

\subsection{Acoustic Side-Channel Features for Verification}
Acoustic side-channel verification does not require any changes to the deployed environment. As a completely passive approach, a floor manager can leave a general purpose computing device running our verification techniques, such as a work tablet in the robot’s vicinity to sense and verify robot action. Alternate side-channels such as power consumption and electromagnetic emissions, require additional hardware installation, or direct connection to the system's internal signals. Thus acoustic side-channels are preferable as they are \emph{deployment ready}.

Robotic systems inherently emit distinctive acoustic signals during operation, primarily due to the activity of internal electromechanical components such as motors, actuators, and mechanical joints. These emissions vary based on factors including the movement axis (X, Y, Z), speed, and interactions with the surrounding environment. In this study, we treat these naturally occurring acoustic signals as a side-channel source for behavioral validation, enabling passive monitoring to determine whether the robot performs its assigned tasks as intended.

\subsection{Threat Model}
The integrity of robotic operations depends on the security of the Operating System (OS), associated software, and network stack. We are specifically concerned about active adversaries who have compromised the OS and/or software stack. For instance via software-level manipulation due to poor patch management, including but not limited to unauthorized command injection, control logic tampering, or falsification of actuator or sensor feedback. While the specific method of compromise is not explicitly modelled, the focus is on verifying whether the robot's physical behavior aligns with the intended command, regardless of the integrity of its internal systems.  Through Acoustic Side-Channel Analysis (ASCA), naturally emitted sounds from robotic motion are captured and analyzed to assess whether the robot’s physical execution corresponds to expected behavior. This enables a non-intrusive, external verification strategy that operates independently of the robot’s software-reported state. The primary objective is to detect discrepancies between the issued command and the executed behavior.

\subsection{Hypotheses and Goals}

We investigate the feasibility of using acoustic side-channel signals to verify and differentiate robotic movements during normal operation. We hypothesize that distinct movement characteristics—such as speed, trajectory, and movement direction produce unique acoustic signatures that can be captured and analyzed using time-frequency features, including Doppler shifts, amplitude modulations, and spectral patterns. Furthermore, we examine how recording setup parameters, such as microphone distance and environmental conditions, influence the quality and reliability of the captured signals.

\subsection{Related Work}
\label{sec:related}

Ensuring the safety, reliability, and security of robotic systems has become a critical area of research, particularly in domains such as healthcare, industrial automation, and teleoperation \cite{murphy2000,  aschenbrenner2015teleoperation}. As robotic systems become increasingly networked, they are exposed to a range of cyber-physical threats, including vulnerabilities in communication protocols and control logic \cite{ahmed2020challenges}. These risks underscore the need for robust verification and authentication mechanisms.

Acoustic Side-Channel Analysis (ASCA)~\cite{sokasca} has emerged as a promising technique for profiling robotic behavior based on sound emissions. Prior work has shown that robotic movements
produce distinguishable acoustic signatures, enabling classification
and fingerprinting of actions through passive audio recordings \cite{shah2022fingerprinting}. Earlier foundational work by Genkin et al. \cite{genkin2017} and Asonov et al. \cite{asonov2004} showed that acoustic emissions from electronic devices can inadvertently leak sensitive information. 



While these studies primarily explore ASCA from an adversarial perspective, relatively little attention has been given to its defensive applications. Most existing work focuses on mitigation strategies, such as reducing sound leakage, introducing masking signals, or shielding sensitive components.


In contrast, the use of acoustic side-channel information as a verification mechanism—specifically to confirm whether a robot has executed a given command correctly—remains largely unexplored. This study addresses that gap by proposing a verification framework that leverages passively captured acoustic signals to verify robotic behavior in real time. Our approach is based on the observation that different movements along the X, Y, and Z axes produce unique acoustic signatures. These signatures can be captured and classified using machine learning models trained on labeled data to determine whether the robot's behavior aligns with the issued command, even in the presence of potential system manipulation.

Other side-channel techniques, such as electromagnetic (EM) analysis, have also been explored for extracting sensitive information from computing devices \cite{gandolfi2001ema}.However, these methods often require specialized equipment and are less practical for passive, real-world deployment. In contrast, acoustic monitoring offers a low-cost, non-invasive alternative that is well-suited for safety-critical environments such as surgical and industrial settings.

\section{Methodology}
\label{sec:defence_evaluation}

Our verification models key acoustic signatures for known robotic actions and use them as references to validate real-time behaviour. In the following discussion, we aim to answer the following questions:

\begin{enumerate}[label=(\subscript{R}{{\arabic*}})]
	\item Can a defender fingerprint individual robot movements on each axis, as well as permutations of them?
		  \item How is movement fingerprinting affected by:
  	\begin{enumerate}[label=(\roman*)]
  		\item The speed and distance of movements?
  		\item The distance the recording device (e.g., smartphone) is away from the
            robot?

  	\end{enumerate}
 
  \item Can entire robot workflows be reconstructed from acoustic emanations?

\end{enumerate}

\definecolor{darkgray}{RGB}{80,80,80}

\begin{figure}[h]
	\centering
	\includegraphics[width=1.0\linewidth]{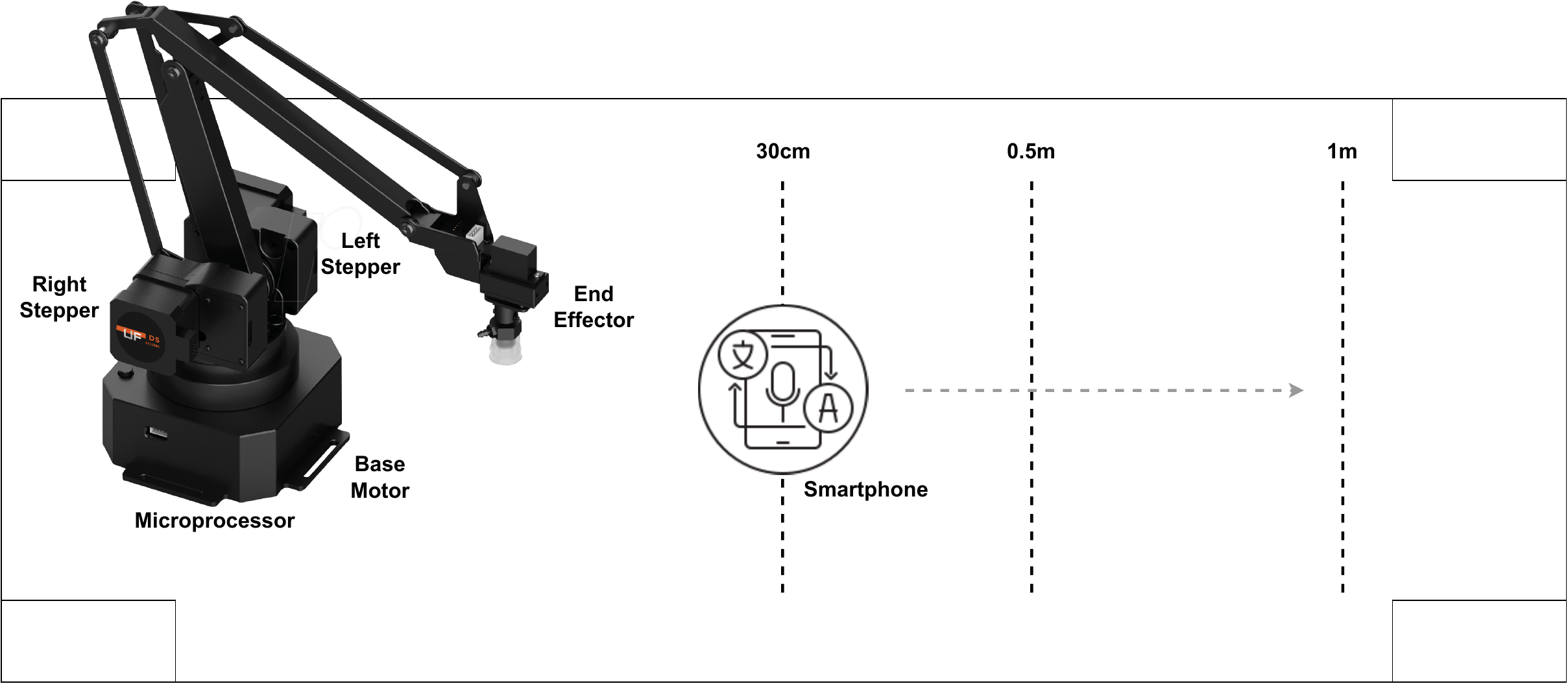}
  \caption{Robot Environment for Acoustic Side Channel. 
 \newline \textcolor{darkgray}{The experimental setup consists of a uArm robot placed on a desktop, performing predefined movements, while a nearby microphone records acoustic emissions for side-channel analysis.} }
  \vspace{-0.5cm}
	\label{fig:robotenv}
\end{figure}

\subsection{Experimental Setup}

This study explores the use of acoustic signals as a passive side-channel to verify robotic behaviour under various operating conditions. The experimental platform is based on the uFactory UArm Swift Pro robotic arm, controlled by an Arduino Mega 2560 running MicroPython. The robot receives commands from a Windows 10 workstation using the uArm Python SDK, which translates high-level instructions into G-code understood by the robotic arm. To capture acoustic emissions generated during robot operation, a smartphone microphone was positioned at changing distances (ranging from 30 cm to 1 m) from the robot. These emissions originate from internal electromechanical activity, including actuator movements, stepper motor operations, and structural vibrations. The experimental setup is designed to reflect realistic deployment scenarios, such as medical laboratories, automated warehouses, and remote-controlled field robotic environments.

A general view of the robot’s operational environment is illustrated in Figure~\ref{fig:robotenv}. The robot is placed centrally on a worktable, and the recording device is positioned at various distances to capture acoustic signals during task execution. This configuration enables the study of how spatial and environmental factors influence the fidelity and reliability of acoustic-based behavioural verification.

\subsection{Acoustic Verification Features}
\label{acoustic_features}
To prepare the audio data for verification, raw acoustic signals are captured using a passive microphone (e.g., a smartphone or external microphone) and subsequently preprocessed through bandpass filtering and normalisation. The audio stream is segmented into short, overlapping frames—typically 20–40 milliseconds in duration with 50\% overlap—to capture temporal and spectral variations throughout the robot’s motion. From each frame, a feature vector is extracted to represent the movement's unique acoustic signature. The following key features are utilised for behavioural classification:

\paragraphb{Root Mean Square Energy (RMSE).} RMSE reflects the signal's energy over time and serves as an indicator of motion intensity. Louder or faster robotic actions typically generate higher energy values, making this feature helpful in differentiating between movement types \cite{panagiotakis2005speech}.

\paragraphb{Zero-Crossing Rate (ZCR).} ZCR counts the number of times the signal crosses the zero amplitude axis within a frame. It is sensitive to sharp or noisy signals and helps identify vibration characteristics linked to rapid mechanical changes \cite{panagiotakis2005speech}.

\paragraphb{Spectral Centroid.}
This feature represents the "center of gravity" of the frequency spectrum and is associated with the perceived brightness or sharpness of a sound. Faster or more forceful movements tend to have higher centroid values \cite{le2011spectral}.

\paragraphb{Spectral Bandwidth.}
Bandwidth describes the spread of the frequency components around the centroid. It helps to differentiate between simple, narrow-band signals and more complex multi-component sounds \cite{atahan2021music, klapuri2006signal}.

\paragraphb{Spectral Rolloff.} This feature identifies the frequency below which a fixed percentage (e.g., 85\%) of the total spectral energy resides. It is useful in separating harmonic movement signals from background or wideband noise \cite{kos2013acoustic}.

\paragraphb{Spectral Contrast.} Spectral contrast captures the difference in energy between spectral peaks and valleys. It reflects how tones or frequencies change over time and is especially relevant for movements that involve gear changes or multi-axis motion \cite{jiang2002music}.

\paragraphb{Chroma Feature.} These features map energy across the 12 pitch classes (C through B) and are typically used to characterise tonal and harmonic content \cite{cho2014relative, paulus2010state}. In this study, they were included as candidate features for capturing tonal variations in robot-generated audio.

\paragraphb{Mel-Frequency Cepstral Coefficients (MFCCs).} MFCCs are derived from the Mel scale, a perceptual pitch scale designed to approximate how humans perceive differences in frequency \cite{greenwood1997mel}. In acoustic signal processing, MFCCs represent the short-term power spectrum of a sound and are widely used in speech, and mechanical system analysis \cite{martinez2012speaker}.
For robotic systems, MFCCs provide a compact representation of the spectral shape of movement-generated acoustic emissions. These coefficients capture tonal variations resulting from differences in movement axis, motor speed, or actuator behaviour. Since MFCCs model human sound perception, they are particularly robust to environmental variation and background noise when processed with an audio-processing pipeline designed to emulate human hearing. MFCCs were used as primary features in classifying robotic motion. Each distinct robot action produces a unique MFCC signature, which can be compared to a baseline to verify whether the commanded behaviour was executed correctly.

\subsection{Operational Parameters and Evaluation Conditions}

To evaluate the applicability of acoustic side-channel signals for verifying robotic behaviour, experiments were conducted across a range of operational and environmental conditions. These parameters were systematically varied to assess their influence on the accuracy and robustness of the acoustic classification. Specifically, we examine the effects of robot movement speed and distance, and microphone placement.

\paragraphb{Speed and Distance.}
The robot was programmed to execute movement commands along the X, Y, and Z axes, with variations in both the movement distance (measured in millimetres) and the movement speed (in mm/s). These variations are intended to reflect realistic remote operation scenarios in which robotic movement is rarely uniform. By analysing precisely controlled movements under dynamic conditions, we assess whether the acoustic verification model can reliably distinguish between subtle differences in trajectory and momentum.

\paragraphb{Microphone Placement.}
The placement of the recording device relative to the robot is a critical factor in determining the quality of the captured acoustic signals. As the acoustic intensity diminishes with distance, the verification system must be resilient to signal distortion. In our experiments, a smartphone or external microphone was placed at distances ranging from 30 cm to 1 meter from the robot. This setup simulates practical deployment scenarios in environments such as industrial automation or telesurgery, where close proximity to the robot may be unsafe or impractical. Given the compact size of the uArm Swift Pro robot ($150,\text{mm} \times 140,\text{mm} \times 281,\text{mm}$) and the relatively low amplitude of its acoustic emissions, we limited our evaluation to distances of 1 meter or less. This constraint ensures that the captured signals remain within a usable signal-to-noise ratio for analysis.

Collectively, these parameters allow for a robust assessment of how acoustic-based verification strategies perform in practical deployment environments. The results of these experiments provide insight into the feasibility of using passive acoustic monitoring as a non-invasive, real-time integrity verification mechanism for remotely controlled robotic systems.

\definecolor{dark gray}{RGB}{80,80,80}

\begin{figure}
	\centering
	\includegraphics[width=\linewidth]{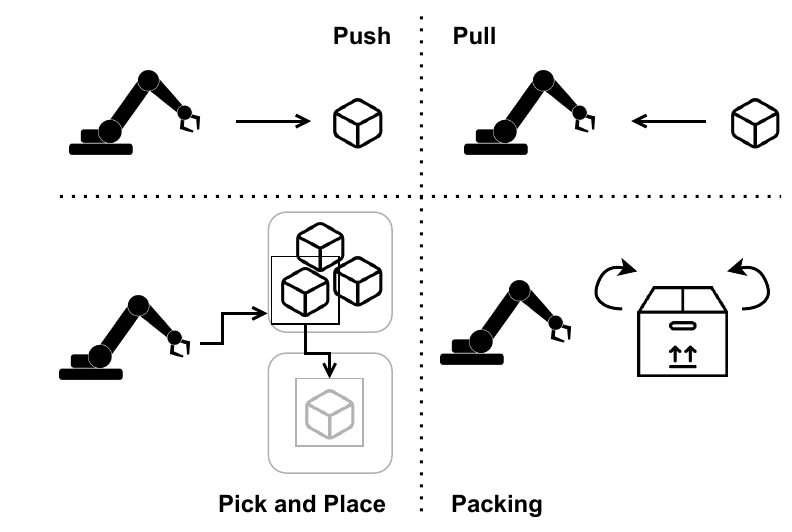}
    \caption{\centering Depiction of Common Warehousing Workflows.}
   \vspace{ 0 cm}
    \textcolor{darkgray}{
    The experimental dataset comprises various robotic movement workflows, including pushing, pulling, packing, and carrying objects, which are utilised to analyse acoustic emissions and validate robot behaviour using machine learning techniques.    }
	\label{fig:workflows}
\end{figure}

\subsection{Movement Dataset}

To develop and evaluate an acoustic-based framework for validating robotic behaviour, we constructed a comprehensive dataset comprising audio recordings of a robotic arm performing a series of controlled movements. These include both individual axis movements (X, Y, Z) and composite workflows that simulate realistic remote-controlled tasks, such as pick-and-place and push-and-pull operations. The dataset consists of two primary subsets. The first subset contains original audio recordings collected under varying operational conditions, including different movement speeds, movement distances, and microphone placements (ranging from 30 cm to 1 m). These variations are designed to reflect the diversity of real-world deployment environments. 

\paragraph{Dataset Preprocessing.}
Feature extraction was performed using the \texttt{librosa} Python library. Each audio recording was segmented into 1-second chunks, using the sampling rate inferred from the original file. For each chunk, a set of time-domain and frequency-domain features was computed to characterize the acoustic signature of the robot's movement. Details on the extracted features are provided in section~\ref{acoustic_features}.

Spectral features and MFCCs were derived from a Short-Time Fourier Transform (STFT) using a Hann window. While the FFT window size was not explicitly specified in the code, it defaults to 2048 in \texttt{librosa}. Future configurations may explore larger window sizes (e.g., 8192) to improve frequency resolution. For MFCCs, 14 coefficients were extracted per frame. Unlike conventional approaches that discard the zeroth coefficient (which represents average log-energy), all 14 coefficients—including the zeroth—were retained in this study. This decision was motivated by the novelty of the application domain, allowing us to explore whether the zeroth coefficient contributes to meaningful distinctions in acoustic fingerprinting of robotic movement. Each audio chunk was annotated with fixed metadata, including movement type, microphone distance, speed, and movement distance. This metadata was used to infer the relationship between acoustic features and robot movements. The extracted features were used to train classification models to predict the most likely movement type and speed class from the observed acoustic signature.

\subsection{Machine Learning (ML) Methods}

In order to evaluate the applicability of the acoustic-based fingerprint approach in verifying robotic movements, various machine and deep learning-based models based on the extracted audio data were designed and implemented. In this context, four basic models are examined: Convolutional Neural Network (CNN), Recurrent Neural Network (RNN), Deep Neural Network (DNN), and Support Vector Machine (SVM) as a comparative non-neural basis. All neural network-based models were implemented with the TensorFlow backend using the Keras API and trained using the Adam optimization algorithm. Depending on how the labels are encoded, either categorical or sparse categorical cross-entropy is used as the loss function. This framework aims to systematically compare the extent to which different deep learning architectures can distinguish movement patterns derived from acoustic features.

\paragraphb{Convolutional Neural Network (CNN):}
Before the evaluation phase, it is necessary to design a suitable neural network architecture that effectively learns discriminative patterns from the input data. For this purpose, a sequential convolutional neural network (CNN) is implemented using the Keras API. The CNN approach automatically captures local relationships and invariant patterns in task-oriented datasets, providing more generalizable results across different tasks or movement categories. After evaluation, the CNN and 3 other classification methods (SVM, DNN, and RNN) will be compared.

The model architecture consists of convolution, pooling, and regularization layers. Nonlinear transformations (in convolutional layers) are performed using the ReLU activation function, and pooling layers are used for dimensionality reduction and preservation of salient features. Batch normalization and dropout layers were added to increase model stability and prevent overfitting. The dense structure in the last layer produces a probability distribution for each class using the softmax activation function.

The network is trained with a categorical cross-entropy loss function and the Adam optimization algorithm, both of which are suitable for multi-class classification. The training process was monitored using early stopping and reduction in learning rate to ensure optimal weights were maintained. Model performance was evaluated using accuracy and loss curves during the training and validation phases, followed by analysis of precision, recall, F1 score, and the confusion matrix. These results comprehensively demonstrated the model's inter-class discrimination and overall generalization performance.

\paragraphb{Deep Neural Network (DNN):}
A fully connected Deep Neural Network (DNN) model was applied to classify the feature vectors obtained from the dataset. The DNN architecture is used for its ability to learn nonlinear relationships formed by large numbers of features and to extract hierarchical representations from tabular data. The model consists of multiple fully connected (dense) layers; the ReLU activation function is used in these layers, and overfitting is prevented by dropout regularization. In the last layer, the probability distribution over all target movement classes is obtained using the softmax activation function.

The network was trained using the Adam optimization algorithm and a sparse categorical cross-entropy loss function. During training, accuracy and loss metrics were monitored on both training and validation sets. The model's generalization performance was improved with early stopping and regularization, and was analyzed using the classification report (precision, recall, F1-score) and a confusion matrix in the final evaluation.

\paragraphb{Recurrent Neural Network (RNN):}
To capture potential temporal dependencies among acoustic features, a Recurrent Neural Network (RNN) with stacked Long Short-Term Memory (LSTM) layers was used.  Although the input data are not inherently sequential, the extracted feature vectors were reshaped into a time-step format $(\text{samples}, \text{timesteps}, \text{features})$, allowing the model to treat each feature vector as a short sequence.

The architecture consists of two consecutive LSTM layers: The first LSTM layer contains 64 units and is configured to return sequences, enabling stacking with a second LSTM layer of 32 units. Each LSTM block is followed by a dropout layer with a rate of 0.2 to reduce overfitting. The final dense output layer consists of seven neurons with softmax activation for multi-class classification.

This model was trained using sparse categorical cross-entropy and the Adam optimiser over 20 epochs with a batch size of 32. The input shape was dynamically adjusted based on the number of features. The goal was to evaluate whether LSTM-based architectures can effectively capture inter-feature dependencies in a classification context.

\paragraphb{Support Vector Machine (SVM):}
Although SVMs do not benefit from backpropagation or deep learning optimisations, they are well-suited for high-dimensional feature spaces and serve as a strong baseline for structured data classification. SVM classifier with a radial basis function (RBF) kernel is also implemented. Input features are normalised using min-max scaling to eliminate scale differences, and movement labels are encoded as integers. To maintain balanced class distribution during training and evaluation, a stratified training-validation split was used. The SVM was trained with a regularisation parameter $C = 1.0$ and automatic gamma scaling. 

\paragraphb{Activation and Optimization Functions:}
Across all deep learning models, the ReLU activation function was selected for hidden layers due to its computational efficiency and reduced susceptibility to vanishing gradient issues. Softmax activation was used in the output layers to enable multi-class classification by producing normalized probability distributions. The Adam optimizer was used to train the neural-network models.

\begin{table}
  \centering
   \scriptsize
  \begin{tabular}{lcc|cc|cc|ccc}
    \toprule
	\multicolumn{10}{c}{{M = Movement, P = Precision, R = Recall}} \\
	\midrule
    \multirow{2}{*}{} &
      \multicolumn{2}{c}{SVM} &
      \multicolumn{2}{c}{DNN} &
	  \multicolumn{2}{c}{RNN} &
	  \multicolumn{2}{c}{CNN} & \\ M
      & {P} & {R} & {P} & {R} & {P} & {R} & {P} & {R} \\
      \midrule
    X & 98\% & 99\% & 99\% & 99\% & 97\% & 98\% & 98\% & 99\% \\
    Y & 97\% & 99\% & 97\% & 98\% & 98\% & 95\% & 96\% & 99\%  \\
    Z & 92\% & 94\% & 92\% & 95\% & 90\% & 92\% & 95\% & 93\%  \\
    XY & 97\% & 97\% & 98\% & 95\% & 97\% & 94\% & 98\% & 97\% \\
    XZ & 66\% & 65\% & 73\% & 68\% & 72\% & 54\% & 67\% & 74\% \\
    YZ & 58\% & 59\% & 67\% & 61\% & 54\% & 58\% & 66\% & 55\% \\
    XYZ & 67\% & 67\% & 69\% & 80\% & 60\% & 72\% & 74\% & 78\% \\
    \midrule
    Accuracy & \multicolumn{2}{c}{83\%} & \multicolumn{2}{c}{85\%} & \multicolumn{2}{c}{80\%} & \multicolumn{2}{c}{85\%} \\
    \bottomrule
  \end{tabular}
  \caption{\centering Baseline Classification Results for SVM, DNN, RNN, and CNN}
  \vspace{-0.5 cm}
  \hspace{\textwidth}{\textcolor{darkgray}{\small\textmd{Overall, the baseline accuracy for DNN and CNN is 85\%, the highest, indicating strong verification ability.  Among the tested movements, the lowest overall precision and recall were observed for the YZ movement. The highest precision and recall are observed for the X and Y movements, though there are exceptions.
  }}}
  \label{table:baseline}
\end{table}

\section{Evaluation}

Following the setup of the robotic environment and the collection of acoustic emissions during various movement commands, we proceed to evaluate the effectiveness of the proposed acoustic-based verification mechanism. The evaluation is structured according to the research questions outlined earlier.

\subsection{Individual Movement Fingerprints}

The first research question ($R_1$) investigates whether individual robot movements along the X, Y, and Z axes, as well as their combinations, can be accurately verified using acoustic signals captured during execution. This experiment serves as a baseline scenario, in which the robot performs movements at a minimum speed of 12.5 mm/s and a minimum movement distance of 1 mm.

As shown in Table~\ref{table:baseline}, the classification model achieves an average accuracy of approximately 80\% across all movement types. Among the individual axes, precision and recall values are generally comparable between X and Y movements, with X-axis achieving marginally higher precision in most models. Notably, movements involving the YZ axis exhibit a significant disparity between precision and recall, suggesting potential confusion in distinguishing these combined movements.

Additionally, movements along the Z-axis consistently result in lower classification performance. This may be attributed to the nature of Z-axis movement, which is restricted to vertical displacement and does not produce significant variation in acoustic characteristics relative to the microphone’s position. Unlike X or Y movements, Z-axis actions do not approach or recede from the microphone, potentially limiting the richness of the captured acoustic signature.

Furthermore, comparing all methods, the highest accuracy is observed for CNN and DNN, at 85\%, while the lowest accuracy is achieved by RNN, at 80\%. SVM, on the other hand, is very close to the highest value, but has an average accuracy of 83\%. Surprisingly, although the precision and recall values for CNN and DNN differ, their accuracies are the same.

\begin{table}
  \centering
   \scriptsize
  \begin{tabular}{lcc|cc|cc|cc|ccc}
    \toprule
	\multicolumn{12}{c}{{D = Distance (mm), P = Precision, R = Recall}} \\
	\midrule
    \multirow{2}{*}{} &
      \multicolumn{2}{c}{D=2} &
      \multicolumn{2}{c}{D=5} &
	  \multicolumn{2}{c}{D=10} &
	  \multicolumn{2}{c}{D=25} &
	  \multicolumn{2}{c}{D=50} & \\
      & {P} & {R} & {P} & {R} & {P} & {R} & {P} & {R} & {P} & {R} \\
      \midrule
    X & 84\% & 89\% & 81\% & 93\% & 94\% & 98\% & 96\% & 95\% & 93\% & 98\% \\
    Y & 92\% & 93\% & 92\% & 90\% & 85\% & 87\% & 87\% & 82\% & 95\% & 95\% \\
    Z & 78\% & 60\% & 92\% & 85\% & 81\% & 88\% & 81\% & 88\% & 97\% & 96\% \\
    XY & 91\% & 94\% & 94\% & 94\% & 84\% & 84\% & 83\% & 89\% & 92\% & 91\% \\
    XZ & 87\% & 87\% & 90\% & 87\% & 91\% & 87\% & 79\% & 79\% & 95\% & 94\% \\
    YZ & 95\% & 90\% & 96\% & 97\% & 92\% & 81\% & 81\% & 78\% & 89\% & 86\% \\
    XYZ & 85\% & 90\% & 94\% & 92\% & 84\% & 87\% & 84\% & 78\% & 82\% & 83\% \\
    \midrule
    Accuracy & \multicolumn{2}{c}{89\%} & \multicolumn{2}{c}{91\%} & \multicolumn{2}{c}{87\%} & \multicolumn{2}{c}{84\%} & \multicolumn{2}{c}{92\%} \\
    \bottomrule
  \end{tabular}
  \caption{\centering Classification Results for Distance Moved (SVM)}
  \vspace{-0.5 cm}
  \hspace{\textwidth}{\textcolor{darkgray}{\small\textmd{Except D=25 mm, there is not much variation in the accuracy with respect to distance moved. This indicates that the acoustic fingerprints are preserved irrespective of the distance moved. }}}
  \label{table:acdistancesvm}
\end{table}

\begin{table}
  \centering
   \scriptsize
  \begin{tabular}{lcc|cc|cc|cc|ccc}
    \toprule
	\multicolumn{12}{c}{{D = Distance (mm), P = Precision, R = Recall}} \\
	\midrule
    \multirow{2}{*}{} &
      \multicolumn{2}{c}{D=2} &
      \multicolumn{2}{c}{D=5} &
	  \multicolumn{2}{c}{D=10} &
	  \multicolumn{2}{c}{D=25} &
	  \multicolumn{2}{c}{D=50} & \\
      & {P} & {R} & {P} & {R} & {P} & {R} & {P} & {R} & {P} & {R} \\
      \midrule
    X & 90\% & 85\% & 87\% & 92\% & 94\% & 97\% & 93\% & 95\% & 95\% & 96\% \\
    Y & 86\% & 96\% & 93\% & 94\% & 85\% & 83\% & 88\% & 84\% & 97\% & 94\% \\
    Z & 77\% & 67\% & 95\% & 86\% & 83\% & 90\% & 78\% & 94\% & 98\% & 96\% \\
    XY & 95\% & 88\% & 95\% & 94\% & 84\% & 84\% & 83\% & 88\% & 95\% & 94\% \\
    XZ & 86\% & 90\% & 92\% & 84\% & 87\% & 92\% & 80\% & 82\% & 95\% & 92\% \\
    YZ & 96\% & 92\% & 93\% & 98\% & 92\% & 81\% & 86\% & 72\% & 88\% & 84\% \\
    XYZ & 87\% & 92\% & 88\% & 94\% & 89\% & 85\% & 82\% & 74\% & 76\% & 85\% \\
    \midrule
    Accuracy & \multicolumn{2}{c}{89\%} & \multicolumn{2}{c}{92\%} & \multicolumn{2}{c}{88\%} & \multicolumn{2}{c}{84\%} & \multicolumn{2}{c}{91\%} \\
    \bottomrule
  \end{tabular}
  \caption{\centering Classification Results for Distance Moved (DNN)}
  \vspace{-0.5 cm}
  \hspace{\textwidth}{\textcolor{darkgray}{\small\textmd{  As with SVM, a small increase in distance generally maintains accuracy, and results are in line with the observation made in Table~\ref{table:acdistancesvm}.
  }}}
  \label{table:acdistancednn}
\end{table}

\begin{table}
  \centering
   \scriptsize
  \begin{tabular}{lcc|cc|cc|cc|ccc}
    \toprule
	\multicolumn{12}{c}{{D = Distance (mm), P = Precision, R = Recall}} \\
	\midrule
    \multirow{2}{*}{} &
      \multicolumn{2}{c}{D=2} &
      \multicolumn{2}{c}{D=5} &
	  \multicolumn{2}{c}{D=10} &
	  \multicolumn{2}{c}{D=25} &
	  \multicolumn{2}{c}{D=50} & \\
      & {P} & {R} & {P} & {R} & {P} & {R} & {P} & {R} & {P} & {R} \\
      \midrule
    X & 81\% & 80\% & 79\% & 84\% & 91\% & 96\% & 89\% & 93\% & 93\% & 95\% \\
    Y & 89\% & 92\% & 86\% & 93\% & 75\% & 77\% & 91\% & 66\% & 90\% & 94\% \\
    Z & 65\% & 31\% & 84\% & 89\% & 78\% & 85\% & 75\% & 90\% & 94\% & 97\% \\
    XY & 93\% & 89\% & 93\% & 90\% & 75\% & 83\% & 67\% & 91\% & 91\% & 86\% \\
    XZ & 75\% & 86\% & 90\% & 78\% & 92\% & 82\% & 73\% & 78\% & 85\% & 96\% \\
    YZ & 92\% & 87\% & 93\% & 94\% & 88\% & 73\% & 73\% & 66\% & 77\% & 84\% \\
    XYZ & 76\% & 86\% & 92\% & 87\% & 83\% & 82\% & 81\% & 57\% & 86\% & 62\% \\
    \midrule
    Accuracy & \multicolumn{2}{c}{83\%} & \multicolumn{2}{c}{88\%} & \multicolumn{2}{c}{83\%} & \multicolumn{2}{c}{76\%} & \multicolumn{2}{c}{88\%} \\
    \bottomrule
  \end{tabular}
  \caption{\centering Classification Results for Distance Moved (RNN)}
  \vspace{-0.5 cm}
  \hspace{\textwidth}{\textcolor{darkgray}{\small\textmd{As with SVM and DNN, the relationship between the movement distance parameter and the resulting result is similar, but the decreases and increases are more dramatic. The accuracy of RNN is the lowest among SVM, DNN, and CNN. }}}
  \label{table:acdistancernn}
\end{table}

\begin{table}
  \centering
   \scriptsize
  \begin{tabular}{lcc|cc|cc|cc|ccc}
    \toprule
	\multicolumn{12}{c}{{D = Distance (mm), P = Precision, R = Recall}} \\
	\midrule
    \multirow{2}{*}{} &
      \multicolumn{2}{c}{D=2} &
      \multicolumn{2}{c}{D=5} &
	  \multicolumn{2}{c}{D=10} &
	  \multicolumn{2}{c}{D=25} &
	  \multicolumn{2}{c}{D=50} & \\
      & {P} & {R} & {P} & {R} & {P} & {R} & {P} & {R} & {P} & {R} \\
      \midrule
    X & 86\% & 89\% & 83\% & 93\% & 95\% & 96\% & 94\% & 95\% & 95\% & 98\% \\
    Y & 93\% & 90\% & 92\% & 90\% & 86\% & 88\% & 89\% & 85\% & 94\% & 96\% \\
    Z & 58\% & 84\% & 89\% & 87\% & 81\% & 91\% & 77\% & 90\% & 97\% & 97\% \\
    XY & 91\% & 93\% & 94\% & 95\% & 88\% & 88\% & 83\% & 91\% & 95\% & 92\% \\
    XZ & 93\% & 82\% & 91\% & 85\% & 93\% & 87\% & 79\% & 75\% & 95\% & 97\% \\
    YZ & 95\% & 88\% & 96\% & 97\% & 90\% & 84\% & 85\% & 75\% & 92\% & 86\% \\
    XYZ & 86\% & 84\% & 94\% & 92\% & 88\% & 86\% & 83\% & 79\% & 83\% & 86\% \\
    \midrule
    Accuracy & \multicolumn{2}{c}{87\%} & \multicolumn{2}{c}{91\%} & \multicolumn{2}{c}{88\%} & \multicolumn{2}{c}{84\%} & \multicolumn{2}{c}{93\%} \\
    \bottomrule
  \end{tabular}
  \caption{\centering Classification Results for Distance Moved (CNN)}
  \vspace{-0.5 cm}
  \hspace{\textwidth}{\textcolor{darkgray}{\small\textmd{
  As in SVM and DNN, the relationship between changes in the movement distance parameter and results is similar; the relationship between movement distance and accuracy is non-monotonic, with the highest CNN accuracy observed at 50 mm. CNN achieves consistently strong performance across movement distances, reaching its highest accuracy of 93\% at 50 mm. For example, for D=2 mm, SVM and DNN achieve the same highest accuracy (89\%), while CNN has an average accuracy (87\%) and is close to the highest values.
}}}
  \label{table:acdistancecnn}
\end{table}

\subsection{Impact of Movement Distance}

The second research question ($R_2$) examines how two factors influence the effectiveness of acoustic-based behaviour verification: (i) the distance and speed of robot movements ($R_{2(i)}$), and (ii) the distance between the recording device and the robot ($R_{2(ii)}$). This subsection focuses on the first factor—movement distance. As the robot moves farther, it is expected to produce more acoustic data, thereby improving discrimination between different types of movement.

As shown in Table~\ref{table:acdistancesvm},  Table~\ref{table:acdistancednn}, Table~\ref{table:acdistancernn}, and Table~\ref{table:acdistancecnn}, overall classification accuracy remains relatively high across different movement distances, although performance decreases at the 25\,mm condition. Most configurations achieve accuracy above 80\%, although RNN drops to 76\% at 25 mm. Precision and recall also tend to remain high across movement distances, although the trend varies across movement classes. At a movement distance of 5 mm, accuracy increases compared with the 2 mm condition. At 10 mm, although overall accuracy remains high, it decreases slightly.

Peak
accuracy occurred at different distances depending on the model:
DNN peaked at 5 mm, while SVM and CNN achieved their highest
accuracy at 50 mm. At 25 mm, the models experience another de-
crease in accuracy—the lowest observed across all distances—but
performance remains strong. This decrease may be due to acoustic aliasing or signal saturation effects during longer movements. However, X-axis classification remains strong. Finally, at 50 mm, overall accuracy increases at a certain rate compared to the initial level.

Across all models, the RNN has the lowest classification accuracy, while the DNN and CNN are very close to each other and have the highest validation accuracies; SVM also achieved high accuracy. The trend between accuracy and increasing movement distance is similar across all models (an increase-decrease relationship). In other words, when D increases from 2 mm to 5 mm, verification accuracy increases across all classification methods. For example, the verification accuracy for DNN, which was 89\% at 2 mm, became 92\% at 5 mm. After 5 mm, accuracy generally decreased at 10 mm and 25 mm before increasing again at 50 mm, although the exact trend varied across classifiers.

\subsection{Impact of Movement Speed}

The next parameter evaluated in the validation framework is the speed at which the robot moves along each axis ($R_{2(i)}$). As shown in Table~\ref{table:acspeedsvm}, Table~\ref{table:acspeeddnn}, Table~\ref{table:acspeedrnn}, and Table~\ref{table:acspeedcnn} changes in movement speed exhibit a weaker correlation with classification accuracy compared to movement distance. On average, classification performance decreases by approximately 10\% relative to the distance-based evaluations. Interestingly, despite some fluctuations, there is a general upward trend in validation accuracy as speed increases. Most movement classes—excluding a few outliers—show improvements in both precision and recall at higher speeds, although minor drops are observed at certain speed levels.

This trend is noteworthy, as it was initially hypothesized that higher speeds would generate higher-frequency acoustic emissions, thereby producing more distinctive spectral features for classification. The experimental results partially support this hypothesis: while higher speeds often enhance classification performance, the relationship is not strictly linear or consistent across all movement types. These findings suggest that movement speed is an informative dimension in acoustic side-channel analysis and can contribute meaningfully to behavioural verification. With classification accuracy consistently exceeding 70\% across speed variations, the results underscore the importance of incorporating diverse movement parameters into the verification model. When comparing all models, although the RNN has the lowest accuracy in this experiment, the CNN and DNN have the highest. SVM also has average accuracy values in this experiment. However, the lowest verification accuracy is 70\% (corresponding to the lowest speed in the RNN), and the highest is 86\% (corresponding to the highest speed in the DNN).

\begin{table}
  \centering
   \scriptsize
  \begin{tabular}{lcc|cc|cc|ccc}
    \toprule
	\multicolumn{9}{c}{{S = Speed (mm/s), P = Precision, R = Recall}} \\
	\midrule
    \multirow{2}{*}{} &
      \multicolumn{2}{c}{S=25} &
      \multicolumn{2}{c}{S=50} &
	  \multicolumn{2}{c}{S=75} &
	  \multicolumn{2}{c}{S=100} & \\
      & {P} & {R} & {P} & {R} & {P} & {R} & {P} & {R} \\
      \midrule
    X & 85\% & 74\% & 85\% & 87\% & 84\% & 88\% & 90\% & 84\%\\
    Y & 88\% & 88\% & 87\% & 87\% & 80\% & 87\% & 87\% & 88\% \\
    Z & 60\% & 75\% & 59\% & 73\% & 83\% & 86\% & 81\%& 82\%\\
    XY & 86\% & 87\% & 85\% & 87\% & 87\% & 87\% & 94\% & 94\%\\
    XZ & 59\% & 61\% & 59\% & 61\% & 88\% & 85\% & 75\% & 86\%\\
    YZ & 79\% & 67\% & 77\% & 65\% & 72\% & 76\% & 74\% & 72\%\\
    XYZ & 76\% & 77\% & 76\% & 77\% & 82\% & 68\% & 79\% & 72\%\\
    \midrule
    Accuracy & \multicolumn{2}{c}{75\%} & \multicolumn{2}{c}{74\%} & \multicolumn{2}{c}{82\%} & \multicolumn{2}{c}{83\%} \\
    \bottomrule
  \end{tabular}
  \caption{\centering Classification Results With Speed Parameter with SVM}
  \vspace{-0.5 cm}
 \hspace{\textwidth}{\textcolor{darkgray}{\small\textmd{The speed parameter performs worse than the distance parameter on the acoustic side channel; there is a significant difference in verification accuracy, and accuracy generally increases with increasing speed.}}}
  \label{table:acspeedsvm}
\end{table}

\begin{table}
  \centering
   \scriptsize
  \begin{tabular}{lcc|cc|cc|ccc}
    \toprule
	\multicolumn{9}{c}{{S = Speed (mm/s), P = Precision, R = Recall}} \\
	\midrule
    \multirow{2}{*}{} &
      \multicolumn{2}{c}{S=25} &
      \multicolumn{2}{c}{S=50} &
	  \multicolumn{2}{c}{S=75} &
	  \multicolumn{2}{c}{S=100} & \\
      & {P} & {R} & {P} & {R} & {P} & {R} & {P} & {R} \\
      \midrule
    X & 82\% & 80\% & 82\% & 89\% & 82\% & 92\% & 96\% & 91\%\\
    Y & 89\% & 90\% & 94\% & 90\% & 89\% & 84\% & 87\% & 93\% \\
    Z & 65\% & 79\% & 69\% & 81\% & 88\% & 84\% & 85\%& 84\%\\
    XY & 83\% & 96\% & 92\% & 92\% & 87\% & 90\% & 97\% & 94\%\\
    XZ & 76\% & 62\% & 68\% & 71\% & 91\% & 77\% & 86\% & 84\%\\
    YZ & 76\% & 70\% & 72\% & 71\% & 71\% & 77\% & 75\% & 72\%\\
    XYZ & 81\% & 76\% & 85\% & 66\% & 77\% & 79\% & 75\% & 81\%\\
    \midrule
    Accuracy & \multicolumn{2}{c}{79\%} & \multicolumn{2}{c}{80\%} & \multicolumn{2}{c}{83\%} & \multicolumn{2}{c}{86\%} \\
    \bottomrule
  \end{tabular}
  \caption{\centering Classification Results With Speed Parameter with DNN}
  \vspace{-0.5 cm}
 \hspace{\textwidth}{\textcolor{darkgray}{\small\textmd{The accuracy of the verification analysis performed with DNN is higher than that of SVM. The speed parameter performs worse than the distance parameter. Unlike the SVM verification results, the accuracy increases continuously with speed (a decrease is observed at 50 mm/s in SVM). }}}
  \label{table:acspeeddnn}
\end{table}

\begin{table}
  \centering
   \scriptsize
  \begin{tabular}{lcc|cc|cc|ccc}
    \toprule
	\multicolumn{9}{c}{{S = Speed (mm/s), P = Precision, R = Recall}} \\
	\midrule
    \multirow{2}{*}{} &
      \multicolumn{2}{c}{S=25} &
      \multicolumn{2}{c}{S=50} &
	  \multicolumn{2}{c}{S=75} &
	  \multicolumn{2}{c}{S=100} & \\
      & {P} & {R} & {P} & {R} & {P} & {R} & {P} & {R} \\
      \midrule
    X & 76\% & 74\% & 77\% & 74\% & 76\% & 81\% & 83\% & 90\%\\
    Y & 83\% & 86\% & 83\% & 86\% & 75\% & 85\% & 89\% & 82\% \\
    Z & 60\% & 63\% & 60\% & 63\% & 68\% & 85\% & 82\%& 76\%\\
    XY & 89\% & 79\% & 89\% & 79\% & 86\% & 67\% & 83\% & 95\%\\
    XZ & 56\% & 57\% & 56\% & 57\% & 79\% & 79\% & 67\% & 85\%\\
    YZ & 69\% & 52\% & 69\% & 52\% & 60\% & 64\% & 68\% & 51\%\\
    XYZ & 60\% & 78\% & 60\% & 78\% & 80\% & 57\% & 70\% & 62\%\\
    \midrule
    Accuracy & \multicolumn{2}{c}{70\%} & \multicolumn{2}{c}{70\%} & \multicolumn{2}{c}{74\%} & \multicolumn{2}{c}{77\%} \\
    \bottomrule
  \end{tabular}
  \caption{\centering Classification Results With Speed Parameter with RNN}
  \vspace{-0.5 cm}
 \hspace{\textwidth}{\textcolor{darkgray}{\small\textmd{The accuracy of the validation analysis performed with RNN is the lowest among the analysis methods. Unlike the SVM and DNN validation results, the accuracy remains the same at 50 mm/s, and, interestingly, all precision-recall values are the same, except for the X-axis precision value (for S = 25 and 50 mm/s).}}}
  \label{table:acspeedrnn}
\end{table}

\begin{table}
  \centering
   \scriptsize
  \begin{tabular}{lcc|cc|cc|ccc}
    \toprule
	\multicolumn{9}{c}{{S = Speed (mm/s), P = Precision, R = Recall}} \\
	\midrule
    \multirow{2}{*}{} &
      \multicolumn{2}{c}{S=25} &
      \multicolumn{2}{c}{S=50} &
	  \multicolumn{2}{c}{S=75} &
	  \multicolumn{2}{c}{S=100} & \\
      & {P} & {R} & {P} & {R} & {P} & {R} & {P} & {R} \\
      \midrule
    X & 90\% & 76\% & 85\% & 77\% & 88\% & 83\% & 89\% & 89\%\\
    Y & 86\% & 93\% & 87\% & 94\% & 79\% & 89\% & 86\% & 90\% \\
    Z & 67\% & 84\% & 62\% & 85\% & 84\% & 88\% & 85\%& 84\%\\
    XY & 92\% & 89\% & 94\% & 89\% & 90\% & 85\% & 95\% & 94\%\\
    XZ & 65\% & 65\% & 64\% & 59\% & 86\% & 86\% & 74\% & 88\%\\
    YZ & 84\% & 73\% & 87\% & 70\% & 77\% & 75\% & 79\% & 67\%\\
    XYZ & 81\% & 79\% & 80\% & 77\% & 77\% & 76\% & 80\% & 74\%\\
    \midrule
    Accuracy & \multicolumn{2}{c}{80\%} & \multicolumn{2}{c}{79\%} & \multicolumn{2}{c}{83\%} & \multicolumn{2}{c}{84\%} \\
    \bottomrule
  \end{tabular}
  \caption{\centering Classification Results With Speed Parameter with CNN}
  \vspace{-0.5 cm}
 \hspace{\textwidth}{\textcolor{darkgray}{\small\textmd{The acoustic side channel speed parameter experiment demonstrates high accuracy compared to all analysis methods. CNN showed the highest accuracy at the lowest speed, while DNN showed the highest overall accuracy. Surprisingly, accuracy decreased slightly with increasing speed for s = 50 mm/s, but accuracy increased with increasing speed beyond this value.}}}
  \label{table:acspeedcnn}
\end{table}

\subsection{Microphone Distance}
\label{sec:micdist}

In an acoustic side-channel validation framework, the distance between the recording device and the robot plays a critical role in determining classification performance. Since sound intensity decreases with distance (governed by the inverse-square law and affected by propagation effects such as attenuation and environmental noise), a greater distance is expected to degrade the quality of captured acoustic features and therefore reduce the reliability of motion verification. To evaluate this effect, the system was tested through experiments by recording sound at three distances: 30 cm (starting point), 50 cm, and 100 cm. Although these distances are short, they reflect realistic deployment scenarios for small-scale, remotely controlled systems. They also represent practical use cases for non-invasive monitoring with standard mobile devices, such as smartphones or external microphones, in environmental scenarios where close-range placement is impractical or unsafe.

\begin{table}[]
\centering
\begin{tabular}{lcccccc}
\toprule
\multicolumn{7}{c}{M = Mic Distance (cm), P = Precision, R = Recall}                                                   \\ \midrule
\multicolumn{1}{c}{} & \multicolumn{2}{c|}{M=30}        & \multicolumn{2}{c|}{M=50}        & \multicolumn{2}{c}{M=100} \\
\multicolumn{1}{c}{} & P    & \multicolumn{1}{c|}{R}    & P    & \multicolumn{1}{c|}{R}    & P           & R           \\ \midrule
X                    & 88\% & \multicolumn{1}{c|}{87\%} & 88\% & \multicolumn{1}{c|}{86\%} & 98\%        & 99\%        \\ 
Y                    & 98\% & \multicolumn{1}{c|}{99\%} & 87\% & \multicolumn{1}{c|}{93\%} & 97\%        & 99\%        \\ 
Z                    & 89\% & \multicolumn{1}{c|}{94\%} & 87\% & \multicolumn{1}{c|}{88\%} & 92\%        & 94\%        \\ 
XY                   & 89\% & \multicolumn{1}{c|}{93\%} & 96\% & \multicolumn{1}{c|}{94\%} & 97\%        & 97\%        \\ 
XZ                   & 80\% & \multicolumn{1}{c|}{82\%} & 89\% & \multicolumn{1}{c|}{90\%} & 66\%        & 65\%        \\ 
YZ                   & 70\% & \multicolumn{1}{c|}{72\%} & 88\% & \multicolumn{1}{c|}{85\%} & 58\%        & 59\%        \\ 
XYZ                  & 75\% & \multicolumn{1}{c|}{62\%} & 86\% & \multicolumn{1}{c|}{83\%} & 69\%        & 67\%        \\ \midrule
Accuracy             & \multicolumn{2}{c|}{85\%}         & \multicolumn{2}{c|}{88\%}         & \multicolumn{2}{c}{83\%}  \\ \bottomrule
\end{tabular}
  \caption{\centering Classification Results With Microphone Distance with SVM}
  \vspace{-0.5 cm}
   \hspace{\textwidth}{\textcolor{darkgray}{\small\textmd{As the microphone distance increases, accuracy drops slightly at the highest distance and is highest at the average distance. These results are interesting.}}}
  \label{table:acmicdistsvm}
\end{table}

\begin{table}[]
\centering
\begin{tabular}{lcccccc}
\toprule
\multicolumn{7}{c}{M = Mic Distance (cm), P = Precision, R = Recall}                                                   \\ \midrule
\multicolumn{1}{c}{} & \multicolumn{2}{c|}{M=30}        & \multicolumn{2}{c|}{M=50}        & \multicolumn{2}{c}{M=100} \\
\multicolumn{1}{c}{} & P    & \multicolumn{1}{c|}{R}    & P    & \multicolumn{1}{c|}{R}    & P           & R           \\ \midrule
X                    & 95\% & \multicolumn{1}{c|}{91\%} & 93\% & \multicolumn{1}{c|}{84\%} & 99\%        & 98\%        \\ 
Y                    & 99\% & \multicolumn{1}{c|}{97\%} & 85\% & \multicolumn{1}{c|}{94\%} & 98\%        & 98\%        \\ 
Z                    & 87\% & \multicolumn{1}{c|}{97\%} & 88\% & \multicolumn{1}{c|}{89\%} & 95\%        & 94\%        \\ 
XY                   & 95\% & \multicolumn{1}{c|}{92\%} & 96\% & \multicolumn{1}{c|}{91\%} & 98\%        & 97\%        \\ 
XZ                   & 75\% & \multicolumn{1}{c|}{82\%} & 90\% & \multicolumn{1}{c|}{97\%} & 72\%        & 62\%        \\ 
YZ                   & 72\% & \multicolumn{1}{c|}{79\%} & 90\% & \multicolumn{1}{c|}{88\%} & 66\%        & 52\%        \\ 
XYZ                  & 78\% & \multicolumn{1}{c|}{61\%} & 87\% & \multicolumn{1}{c|}{88\%} & 61\%        & 86\%        \\ \midrule
Accuracy             & \multicolumn{2}{c|}{86\%}         & \multicolumn{2}{c|}{90\%}         & \multicolumn{2}{c}{84\%}  \\ \bottomrule
\end{tabular}
  \caption{\centering Classification Results With Microphone Distance with DNN}
  \vspace{-0.5 cm}
   \hspace{\textwidth}{\textcolor{darkgray}{\small\textmd{As the microphone distance increases, accuracy drops slightly at the highest distance and is highest at the average distance, and these results are interesting.}}}
  \label{table:acmicdistdnn}
\end{table}

\begin{table}[]
\centering
\begin{tabular}{lcccccc}
\toprule
\multicolumn{7}{c}{M = Mic Distance (cm), P = Precision, R = Recall}                                                   \\ \midrule
\multicolumn{1}{c}{} & \multicolumn{2}{c|}{M=30}        & \multicolumn{2}{c|}{M=50}        & \multicolumn{2}{c}{M=100} \\
\multicolumn{1}{c}{} & P    & \multicolumn{1}{c|}{R}    & P    & \multicolumn{1}{c|}{R}    & P           & R           \\ \midrule
X                    & 91\% & \multicolumn{1}{c|}{84\%} & 76\% & \multicolumn{1}{c|}{83\%} & 97\%        & 98\%        \\ 
Y                    & 98\% & \multicolumn{1}{c|}{97\%} & 86\% & \multicolumn{1}{c|}{78\%} & 98\%        & 95\%        \\ 
Z                    & 84\% & \multicolumn{1}{c|}{89\%} & 82\% & \multicolumn{1}{c|}{86\%} & 90\%        & 92\%        \\ 
XY                   & 85\% & \multicolumn{1}{c|}{92\%} & 97\% & \multicolumn{1}{c|}{88\%} & 97\%        & 94\%        \\ 
XZ                   & 75\% & \multicolumn{1}{c|}{74\%} & 86\% & \multicolumn{1}{c|}{93\%} & 72\%        & 54\%        \\ 
YZ                   & 70\% & \multicolumn{1}{c|}{62\%} & 86\% & \multicolumn{1}{c|}{83\%} & 54\%        & 58\%        \\ 
XYZ                  & 63\% & \multicolumn{1}{c|}{68\%} & 84\% & \multicolumn{1}{c|}{84\%} & 60\%        & 72\%        \\ \midrule
Accuracy             & \multicolumn{2}{c|}{81\%}         & \multicolumn{2}{c|}{85\%}         & \multicolumn{2}{c}{80\%}  \\ \bottomrule
\end{tabular}
  \caption{\centering Classification Results With Microphone Distance with RNN}
  \vspace{-0.5 cm}
   \hspace{\textwidth}{\textcolor{darkgray}{\small\textmd{As the distance from the microphone to the recorded robot increases, accuracy drops slightly at the highest distance and is highest at the average distance. These results are interesting, and a similar situation is observed for DNN, SVM, and CNN. RNN again has the lowest accuracy values.
    }}}
  \label{table:acmicdistrnn}
\end{table}

\begin{table}[]
\centering
\begin{tabular}{lcccccc}
\toprule
\multicolumn{7}{c}{M = Mic Distance (cm), P = Precision, R = Recall}                                                   \\ \midrule
\multicolumn{1}{c}{} & \multicolumn{2}{c|}{M=30}        & \multicolumn{2}{c|}{M=50}        & \multicolumn{2}{c}{M=100} \\
\multicolumn{1}{c}{} & P    & \multicolumn{1}{c|}{R}    & P    & \multicolumn{1}{c|}{R}    & P           & R           \\ \midrule
X                    & 92\% & \multicolumn{1}{c|}{94\%} & 94\% & \multicolumn{1}{c|}{91\%} & 97\%        & 99\%        \\ 
Y                    & 98\% & \multicolumn{1}{c|}{99\%} & 90\% & \multicolumn{1}{c|}{95\%} & 97\%        & 98\%        \\ 
Z                    & 92\% & \multicolumn{1}{c|}{95\%} & 87\% & \multicolumn{1}{c|}{95\%} & 94\%        & 97\%        \\ 
XY                   & 92\% & \multicolumn{1}{c|}{95\%} & 98\% & \multicolumn{1}{c|}{93\%} & 98\%        & 96\%        \\ 
XZ                   & 79\% & \multicolumn{1}{c|}{87\%} & 95\% & \multicolumn{1}{c|}{93\%} & 75\%        & 67\%        \\ 
YZ                   & 80\% & \multicolumn{1}{c|}{80\%} & 92\% & \multicolumn{1}{c|}{91\%} & 68\%        & 69\%        \\ 
XYZ                  & 83\% & \multicolumn{1}{c|}{66\%} & 91\% & \multicolumn{1}{c|}{89\%} & 76\%        & 81\%        \\ \midrule
Accuracy             & \multicolumn{2}{c|}{88\%}         & \multicolumn{2}{c|}{92\%}         & \multicolumn{2}{c}{87\%}  \\ \bottomrule
\end{tabular}
  \caption{\centering Classification Results With Microphone Distance with CNN}
  \vspace{-0.5 cm}
   \hspace{\textwidth}{\textcolor{darkgray}{\small\textmd{  
 As the distance between the microphone and the recorded robot increases, the accuracy decreases slightly at the highest distance and is highest at the average distance; this trend also applies to the CNN. Compared to the others, CNN has the highest accuracy values.
 }}}
  \label{table:acmicdistcnn}
\end{table}

As shown in Table~\ref{table:acmicdistsvm}, 
Table~\ref{table:acmicdistdnn},
Table~\ref{table:acmicdistrnn}, and
Table~\ref{table:acmicdistcnn}, verification accuracy generally increased from 30 cm to 50 cm before decreasing at 100 cm, showing that classification performance varied across the evaluated microphone distances. These results suggest that microphone placement and multi-angle acoustic monitoring may play a more nuanced role in system performance than previously assumed. While some accuracy reduction is observed at 100 cm, it is not substantial. This indicates that the system retains a reasonable level of robustness even at extended distances. However, the variability in performance across different movement types also highlights the importance of carefully considering microphone positioning in real-world deployments, as certain configurations may enhance signal clarity while others may introduce noise or ambiguity. Overall, classification accuracy remained high across the evaluated microphone distances, with CNN achieving the highest accuracy and RNN the lowest. Even the lowest observed accuracy remained at 80\%, while the highest exceeded 90\%.

\subsection{Workflow Recovery}
The next step in evaluating our study is to validate different robot movement sequences (pulling, pushing, packing, and pick-and-place) through acoustic analysis. Individual robot movement fingerprints can be recognized with reasonable accuracy, and verifying complete sequences of robot activity can also serve as a powerful tool to ensure operational integrity and detect anomalies. This is particularly important in safety-critical or safety- or security-sensitive environments (such as healthcare or manufacturing) where deviations from expected behavior may indicate malfunctions or environmental interference. As shown in Table~\ref{table:acworkflowrecovery}, our method achieves the maximum accuracy rate of 86\% with DNN in validating movement sequences, while the lowest is 74\% with RNN. Surprisingly, the SVM and CNN values are the same at 84\%, and although this value is close to the highest value, it is lower. Outside of RNN, the precision and recall values for pick-and-place, a complex operation, are low as expected (the lowest precision in RNN is for the push operation). In general, the packaging process has the highest precision and recall values. 

These results indicate that verification by acoustic side-channel analysis is a promising, non-invasive method for monitoring robot behavior involving workflow movements and may contribute to future control or compliance systems for robotic operations.

\begin{table}
  \centering
   \scriptsize
  \begin{tabular}{lcc|cc|cc|ccc}
    \toprule
	\multicolumn{10}{c}{{W = Workflow, P = Precision, R = Recall}} \\
	\midrule
    \multirow{2}{*}{} &
      \multicolumn{2}{c}{SVM} &
      \multicolumn{2}{c}{DNN} &
	  \multicolumn{2}{c}{RNN} &
	  \multicolumn{2}{c}{CNN} & \\ W
      & {P} & {R} & {P} & {R} & {P} & {R} & {P} & {R} \\
      \midrule
    Push & 82\% & 89\% & 84\% & 88\% & 66\% & 86\% & 83\% & 80\% \\
    Pull & 83\% & 87\% & 87\% & 85\% & 82\% & 73\% & 83\% & 86\%  \\
    Pick-and-Place & 79\% & 73\% & 84\% & 81\% & 81\% & 52\% & 82\% & 76\%  \\
    Packing & 93\% & 86\% & 90\% & 92\% & 73\% & 86\% & 88\% & 94\% \\
    \midrule
    Accuracy & \multicolumn{2}{c}{84\%} & \multicolumn{2}{c}{86\%} & \multicolumn{2}{c}{74\%} & \multicolumn{2}{c}{84\%} \\
    \bottomrule
  \end{tabular}
  \caption{\centering Workflow Reconstruction with SVM, DNN, RNN, and CNN}
  \vspace{-0.5 cm}
  \hspace{\textwidth}{\textcolor{darkgray}{\small\textmd{
In the workflow experiment, the DNN achieved the highest accuracy of 86\%, while the RNN achieved the lowest of 74\%. Packing generally showed the highest precision. The most complex movement, pick and place, had the lowest recall and precision (except for the RNN's precision).
  }}}
  \label{table:acworkflowrecovery}
\end{table}
\section{Discussion} 
\label{sec:discussion}

\begin{table}
  \centering
   \scriptsize
  \begin{tabular}{lcc|cc|cc|ccc}
    \toprule
	\multicolumn{10}{c}{{M = Movement, P = Precision, R = Recall}} \\
	\midrule
    \multirow{2}{*}{} &
      \multicolumn{2}{c}{SVM} &
      \multicolumn{2}{c}{DNN} &
	  \multicolumn{2}{c}{RNN} &
	  \multicolumn{2}{c}{CNN} & \\ M
      & {P} & {R} & {P} & {R} & {P} & {R} & {P} & {R} \\
      \midrule
    X & 98\% & 99\% & 99\% & 98\% & 97\% & 98\% & 98\% & 99\% \\
    Y & 98\% & 98\% & 98\% & 98\% & 98\% & 95\% & 97\% & 99\%  \\
    Z & 91\% & 95\% & 94\% & 95\% & 91\% & 91\% & 94\% & 95\%  \\
    XY & 98\% & 97\% & 99\% & 96\% & 96\% & 96\% & 98\% & 98\% \\
    XZ & 66\% & 66\% & 70\% & 68\% & 72\% & 54\% & 65\% & 79\% \\
    YZ & 56\% & 60\% & 64\% & 52\% & 51\% & 58\% & 67\% & 50\% \\
    XYZ & 67\% & 60\% & 64\% & 80\% & 58\% & 65\% & 73\% & 75\% \\
    \midrule
    Accuracy & \multicolumn{2}{c}{82\%} & \multicolumn{2}{c}{84\%} & \multicolumn{2}{c}{79\%} & \multicolumn{2}{c}{85\%} \\
    \bottomrule
  \end{tabular}
  \caption{\centering Amplitude Filtering Classification Results with SVM, DNN, RNN, and CNN}
  \vspace{-0.5 cm}
  \hspace{\textwidth}{\textcolor{darkgray}{\small\textmd{
In general, the amplitude-filtered data obtained with CNN has the highest accuracy (85\%), while the validation accuracy of the filtered data obtained with RNN is the lowest(79\%). The accuracy of amplitude-filtered data obtained with DNN (84\%)is quite close to that of the CNN with the highest accuracy, although not the highest. The accuracy of amplitude-filtered data obtained with SVM is good, although not the highest (82\%).
  }}}
  \label{table:acbaselinefiltered}
\end{table}

\begin{table}
  \centering
   \scriptsize
  \begin{tabular}{lcc|cc|cc|ccc}
    \toprule
	\multicolumn{10}{c}{{M = Movement, P = Precision, R = Recall}} \\
	\midrule
    \multirow{2}{*}{} &
      \multicolumn{2}{c}{SVM} &
      \multicolumn{2}{c}{DNN} &
	  \multicolumn{2}{c}{RNN} &
	  \multicolumn{2}{c}{CNN} & \\ M
      & {P} & {R} & {P} & {R} & {P} & {R} & {P} & {R} \\
      \midrule
    X & 98\% & 99\% & 98\% & 99\% & 98\% & 99\% & 99\% & 99\% \\
    Y & 93\% & 97\% & 96\% & 96\% & 92\% & 94\% & 97\% & 99\%  \\
    Z & 78\% & 94\% & 90\% & 93\% & 80\% & 91\% & 91\% & 93\%  \\
    XY & 93\% & 96\% & 96\% & 96\% & 95\% & 94\% & 94\% & 98\% \\
    XZ & 52\% & 59\% & 50\% & 82\% & 53\% & 61\% & 52\% & 74\% \\
    YZ & 51\% & 46\% & 54\% & 22\% & 43\% & 32\% & 53\% & 34\% \\
    XYZ & 68\% & 46\% & 67\% & 60\% & 61\% & 56\% & 67\% & 57\% \\
    \midrule
    Accuracy & \multicolumn{2}{c}{76\%} & \multicolumn{2}{c}{78\%} & \multicolumn{2}{c}{75\%} & \multicolumn{2}{c}{79\%} \\
    \bottomrule
  \end{tabular}
  \caption{\centering Low Pass Filter  Classification Results with SVM, DNN, RNN, and CNN}
  \vspace{-0.5 cm}
  \hspace{\textwidth}{\textcolor{darkgray}{\small\textmd{
The low-pass filtering validation accuracy appears to drop by approximately 10\% compared to the baseline results and the amplitude filtering validation accuracy. The overall accuracy trend for the low-pass filter is the same as for amplitude filtering, although the accuracy percentages are lower, i.e., the distributions for the highest accuracy (CNN-79\%) and lowest accuracy (RNN-75\%) are the same according to the method.
  }}}
  \label{table:acbaselinefiltered2}
\end{table}

The acoustic side-channel verification system proposed in this study demonstrates the feasibility of non-invasive behavioral verification for robotic systems using passively captured audio signals. While prior research has primarily focused on acoustic side channels as a vector for adversarial attacks, this work reframes them as a defensive asset for real-time integrity monitoring. This section discusses the implications and limitations of the proposed approach.

A certain level of background noise is inevitable when recording acoustic signals in real-world environments, such as university laboratories or industrial settings. Unwanted sounds such as faint human speech, keyboard clicks, HVAC noise, and ambient noise can degrade the performance of machine learning models used for behavioural classification. Although the baseline experiments yielded relatively high classification accuracy, we further investigated whether digital noise filtering techniques could enhance signal quality and improve model performance. Two standard filtering methods were evaluated: amplitude-based filtering and low-pass filtering.

Audio samples were recorded at a sampling rate of 44.1 kHz and 16-bit resolution, yielding 65,536 possible amplitude values per sample. Frequency-domain analysis using the Fast Fourier Transform (FFT) revealed that most of the robot-generated acoustic energy was concentrated below 1 kHz, with a dominant range under 250 Hz. A consistent 60 Hz spike—attributable to electrical hum from power supplies—was observed, along with peaks near 150 Hz and 200 Hz, likely corresponding to motor vibrations and mechanical arm movements.

For amplitude-based filtering, we used PyDub and NumPy to attenuate low-amplitude background noise. This method applied gain reduction to signal regions below a defined decibel threshold (e.g., -30 dB), preserving the dominant acoustic features while suppressing ambient noise. The audio waveform was converted to a NumPy array, filtered, and then reconstructed into a standard \texttt{.wav} file for model input.

Low-pass filtering was implemented using a Butterworth filter via SciPy, with a cutoff frequency set at 1 kHz based on the FFT analysis. This approach effectively attenuated high-frequency noise while preserving the low-frequency components most relevant to robotic motion. Compared to amplitude filtering, the low-pass filter provided smoother waveforms and more consistent suppression of transient noise sources such as keyboard taps and mouse clicks.

The performance of both filtering methods was evaluated using classification accuracy and sensitivity metrics, as summarized in Tables~\ref{table:acbaselinefiltered} and~\ref{table:acbaselinefiltered2}. While both filtering techniques visibly reduced noise components in the frequency spectrum, neither consistently improved classification performance relative to the unfiltered baseline. Between the two methods, amplitude filtering consistently outperformed low-pass filtering, though it remained slightly below baseline accuracy in most cases. For example, although the verification accuracy in amplitude filtering with CNN, which has the highest verification, is 85\%, the verification accuracy of CNN for the low-pass filter is 79\% (the highest value in their tables). Although CNN achieves the highest accuracy for both filtering methods, RNN achieves the lowest. For example, while the accuracy of the amplitude-filtering RNN is 79\%, it is 75\% for the low-pass filtering.

Although the experimental evaluation in this study used a single robotic system, the acoustic properties obtained are not specific to that system. The sounds emitted by the robotic system arise from the actuation and movement dynamics common to most robotic systems. These findings motivate future investigation of whether models trained on one robotic platform can generalise to other robots executing similar workflows. In this context, although some changes in acoustic properties are observed depending on different types of movements, no major differences in basic sound structures are anticipated. It is also acknowledged that model performance may be affected by the size and diversity of the collected dataset. This constitutes an important research direction for future studies in different environments and with larger data sets. Future work aims to validate this assumption by expanding data collection to include different types of robotic systems and evaluating the model's robustness across them.

Sensor-based verification methods are often used to evaluate the consistency of workflows in robotic systems. However, these methods frequently require additional hardware or access to the system's internal signals. An acoustic-based approach, on the other hand, can perform a similar verification process by analyzing the sounds emitted during robot operation, without requiring additional hardware or system intervention. Furthermore, this analysis can be performed using a single microphone (e.g., a smartphone). Thus, acoustic methods complement rather than compete with sensor-based approaches and, when used together, can create a more reliable, layered verification structure.

Future work will explore combining both filtering techniques or incorporating more advanced spectral gating methods, such as adaptive thresholding. Additionally, dynamically adjusting the filtering strategy based on real-time environmental conditions could further improve robustness. Such enhancements would increase the system’s applicability in diverse deployment scenarios, including hospitals, industrial facilities, and collaborative robotics environments.
\section{Conclusion}
\label{sec:conclusion}

This study demonstrates that acoustic emissions can be effectively used as a verification mechanism for validating robotic behavior. By leveraging low-cost recording devices, such as smartphones, the proposed framework enables passive, non-invasive monitoring with high classification accuracy across both individual movements and composite workflows. The system exhibits robust performance under varying operational conditions, including changes in movement speed and microphone placement. Overall, this work establishes acoustic side-channel analysis as a practical and scalable approach for real-time behavioral verification, contributing to enhanced trust and operational transparency in networked robotic systems.

\section{Acknowledgment}
\textbf{This work was supported by UK National Edge AI Hub and PETRAS.}

\bibliographystyle{splncs04}
\bibliography{references}

@book{murphy2000,
  author    = {Murphy, Robin R and Arkin, Ronald C},
  title     = {Introduction to {AI} Robotics},
  publisher = {MIT press},
  year      = {2000}
}

@inproceedings{humphreys2008,
 author    = {Humphreys, Todd E. and Ledvina, Brent M. and Psiaki, Mark L. and O'Hanlon, Brady W. and Kintner, Paul M.},
  title     = {Assessing the Spoofing Threat: Development of a Portable {GPS} Civilian Spoofer},
  booktitle = {Proceedings of the 21st International Technical Meeting of the Satellite Division of The Institute of Navigation (ION GNSS 2008)},
  pages     = {2314--2325},
  year      = {2008},
  address   = {Savannah, GA}
}

@article{genkin2017,
  author  = {D. Genkin and A. Shamir and E. Tromer},
  title   = {Acoustic cryptanalysis},
  journal = {Journal of Cryptology},
  volume  = {30},
  pages   = {392--443},
  year    = {2017},
  doi = {10.1007/s00145-015-9224-2}
}

@inproceedings{asonov2004,
  author    = {D. Asonov and R. Agrawal},
  title     = {Keyboard acoustic emanations},
  booktitle = {IEEE Symposium on Security and Privacy, 2004. Proceedings. 2004},
  year      = {2004},
  pages={3-11},
  doi      = {10.1109/SECPRI.2004.1301311}
}

@article{ahmed2020challenges,
  title={Challenges and opportunities in cyberphysical systems security: A physics-based perspective},
  author={Ahmed, Chuadhry Mujeeb and Zhou, Jianying},
  journal={IEEE Security and Privacy},
  volume={18},
  number={6},
  pages={14--22},
  year={2020},
  publisher={Institute of Electrical and Electronics Engineers (IEEE)},
  doi = {10.1109/MSEC.2020.3002851}
}

@misc{ahmed2024time,
  title={Time constant: Actuator fingerprinting using transient response of device and process in ics},
  author={Ahmed, Chuadhry Mujeeb and Calder, Matthew and Gunawan, Sean and Prakash, Jay and Nagaraja, Shishir and Zhou, Jianying},
  year={2024},
  doi = {10.48550/arXiv.2409.16536}
}

@inproceedings{gandolfi2001ema,
  author    = {Karla Gandolfi and Christophe Mourtel and Francis Olivier},
  title     = {Electromagnetic analysis: Concrete results},
  booktitle = {Proceedings of the Third International Workshop on Cryptographic Hardware and Embedded Systems},
  volume    = {2162},
  pages     = {251--261},
  year      = {2001},
  doi = {10.1007/3-540-44709-1\_21},
  publisher = {Springer Berlin Heidelberg}
}

@article{aschenbrenner2015teleoperation,
  author = {Aschenbrenner, D. and Fritscher, Michael and Sittner, F. and Krauss, M. and Schilling, K.},
  title     = {Teleoperation of an industrial robot in an active production line},
  journal   = {IFAC-PapersOnLine},
  volume    = {48},
  number    = {10},
  pages     = {159--164},
  year      = {2015},
  doi = {10.1016/j.ifacol.2015.08.125},
  publisher = {Elsevier BV}
}

@inproceedings{avila2020teleoperated,
  author    = {Jose Luis Ordoñez Avila and Hector Jimenez and Tania Marquez and Carlos Muñoz and Alberto Max Carrazco and Maria Elena Perdomo and David Miselem and David Nolasco},
  title     = {Study Case: Teleoperated Voice Picking Robots prototype as a logistic solution in Honduras},
  booktitle = {2020 5th International Conference on Control and Robotics Engineering (ICCRE)},
  pages     = {19--24},
  publisher = {IEEE},
  year      = {2020},
  doi = {10.1109/ICCRE49379.2020.9096483}
}

@article{bartos2021automotive,
  author    = {Michal Bartoš and Vladimír Bulej and Martin Bohušík and Ján Stanček and Vitalii Ivanov and Peter Macek},
  title     = {An overview of robot applications in automotive industry},
  journal   = {Transportation Research Procedia},
  volume    = {55},
  pages     = {837--844},
  year      = {2021},
  doi = {10.1016/j.trpro.2021.07.052},
  publisher = {Elsevier BV}
}

@article{grabowski2021teleoperated,
author  = {Grabowski, Andrzej and Jankowski, Jarosław and Wodzyński, Mieszko},
  title   = {Teleoperated Mobile Robot with Two Arms: The Influence of a Human-Machine Interface, VR Training and Operator Age},
  journal = {International Journal of Human-Computer Studies},
  volume  = {156},
  pages   = {102707},
  year    = {2021},
  doi     = {10.1016/j.ijhcs.2021.102707},
  publisher = {Elsevier BV},
}

@article{li2017teleoperation,
  author  = {Chunxu Li and Chenguang Yang and Jian Wan and Andy SK Annamalai and Angelo Cangelosi},
  title   = {Teleoperation control of Baxter robot using Kalman filter-based sensor fusion},
  journal = {Systems Science \& Control Engineering},
  volume  = {5},
  number  = {1},
  pages   = {156--167},
  year    = {2017},
  doi = {10.1080/21642583.2017.1300109},
  publisher = {Informa UK Limited},
}

@article{hannaford2012raven,
  author  = {Blake Hannaford and Jacob Rosen and Diana W. Friedman and Hawkeye King and Phillip Roan and Lei Cheng and Daniel Glozman and Ji Ma and Sina Nia Kosari and Lee White},
  title   = {Raven-II: an Open Platform for Surgical Robotics Research},
  journal = {IEEE Transactions on Biomedical Engineering},
  volume  = {60},
  number  = {4},
  pages   = {954--959},
  year    = {2013},
  doi  = {10.1109/TBME.2012.2228858},
  publisher = {Institute of Electrical and Electronics Engineers (IEEE)},
}

@article{sung2001laparoscopic,
  author  = {Gyung Tak Sung and Inderbir S. Gill},
  title   = {Robotic Laparoscopic Surgery: a Comparison of the Da Vinci and Zeus Systems},
  journal = {Urology},
  volume  = {58},
  number  = {6},
  pages   = {893--898},
  year    = {2001},
  doi = {10.1016/S0090-4295(01)01423-6},
  publisher = {Elsevier BV}
}

@article{tewari2002prostatectomy,
  author  = {Ashutosh Tewari and James Peabody and Richard Sarle and Guruswami Balakrishnan and Ashok Hemal and Alok Shrivastava and Mani Menon},
  title   = {Technique of da Vinci robot-assisted anatomic radical prostatectomy},
  journal = {Urology},
  volume  = {60},
  number  = {4},
  pages   = {569--572},
  year    = {2002},
  doi = {10.1016/S0090-4295(02)01852-6},
  publisher = {Elsevier BV}
}

@incollection{marescaux2006telesurgery,
  author    = {Marescaux, Jacques and Rubino, Francesco},
  title     = {Transcontinental robot-assisted remote telesurgery, feasibility and potential applications},
  booktitle = {Teleophthalmology},
  publisher = {Springer},
  pages     = {261--265},
  year      = {2006},
  doi = {10.1007/3-540-33714-8\_31}
}

@article{okamura2004haptic,
  author  = {A. M. Okamura},
  title   = {Methods for haptic feedback in teleoperated robot-assisted surgery},
  journal = {Industrial robot},
  volume  = {31},
  number  = {6},
  pages   = {499--508},
  year    = {2004},
   doi = {10.1108/01439910410566362},
    publisher = {Emerald}
}

@article{sokasca,
author  = {Wang, Ping and Nagaraja, Shishir and Bourquard, Aurélien and Gao, Haichang and Yan, Jeff},
  title   = {{SoK}: Acoustic Side Channels},
  journal = {ACM Computing Surveys},
  volume  = {58},
  number  = {7},
  year    = {2026},
  doi     = {10.1145/3778350}
}

@article{panagiotakis2005speech,
  author  = {Panagiotakis, Costas and Tziritas, Georgios},
  title   = {A speech/music discriminator based on RMS and zero-crossings},
  journal = {IEEE Transactions on Multimedia},
  volume  = {7},
  number  = {1},
  pages   = {155--166},
  year    = {2005},
  doi = {10.1109/TMM.2004.840604},
  publisher = {Institute of Electrical and Electronics Engineers (IEEE)}
}

@article{le2011spectral,
  author  = {Le, Phu Ngoc and Ambikairajah, Eliathamby and Epps, Julien and Sethu, Vidhyasaharan and Choi, Eric H. C.},
  title   = {Investigation of spectral centroid features for cognitive load classification},
  journal = {Speech Communication},
  volume  = {53},
  number  = {4},
  pages   = {540--551},
  year    = {2011},
  doi = {10.1016/j.specom.2011.01.005},
  publisher = {Elsevier BV}
}

@inproceedings{atahan2021music,
  author    = {Atahan, Yunus and Elbir, Ahmet and Keskin, Abdullah Enes and Kiraz, Osman and Kirval, Bulent and Aydin, Nizamettin},
  title     = {Music Genre Classification Using Acoustic Features and Autoencoders},
  booktitle = {2021 Innovations in Intelligent Systems and Applications Conference (ASYU)},
  pages     = {1--5},
  year      = {2021},
  doi = {10.1109/ASYU52992.2021.9598979},
  publisher = {IEEE}
}

@book{klapuri2006signal,
  editor    = {Klapuri, Anssi and Davy, Manuel},
  title     = {Signal Processing Methods for Music Transcription},
  year      = {2006},
  publisher = {Springer US},
  doi = {10.1007/0-387-32845-9}
}

@article{kos2013acoustic,
  author  = {Kos, Marko and Kačič, Zdravko and Vlaj, Damjan},
  title   = {Acoustic classification and segmentation using modified spectral roll-off and variance-based features},
  journal = {Digital Signal Processing},
  volume  = {23},
  number  = {2},
  pages   = {659--674},
  year    = {2013},
  doi = {10.1016/j.dsp.2012.10.008},
  publisher = {Elsevier BV},
}

@article{cho2014relative,
  author  = {Cho, Taemin and Bello, Juan P.},
  title   = {On the relative importance of individual components of chord recognition systems},
  journal = {IEEE/ACM Transactions on Audio, Speech, and Language Processing},
  volume  = {22},
  number  = {2},
  pages   = {477--492},
  year    = {2014},
  doi = {10.1109/TASLP.2013.2295926},
   publisher = {Institute of Electrical and Electronics Engineers (IEEE)}
}

@article{greenwood1997mel,
  author  = {Greenwood, Donald D.},
  title   = {The {M}el scale's disqualifying bias and a consistency of pitch-difference equisections in 1956 with equal cochlear distances and equal frequency ratios},
  journal = {Hearing Research},
  volume  = {103},
  pages   = {199--224},
  year    = {1997},
  doi = {10.1016/S0378-5955(96)00175-X},
  publisher = {Elsevier BV}
}

@misc{shah2022fingerprinting,
  author        = {Shah, Ryan and Ahmed, Chuadhry Mujeeb and Nagaraja, Shishir},
  title         = {Fingerprinting Robot Movements via Acoustic Side Channel},
  year          = {2022},
  eprint        = {2209.10240},
  archivePrefix = {arXiv},
  url           = {https://arxiv.org/abs/2209.10240},
  doi = {10.48550/arXiv.2209.10240}
}

@inproceedings{jiang2002music,
  author    = {Jiang, Dan-Ning and Lu, Lie and Zhang, Hong-Jiang and Tao, Jian-Hua and Cai, Lian-Hong},
  title     = {Music type classification by spectral contrast feature},
  booktitle = {Proceedings. IEEE International Conference on Multimedia and Expo},
  volume    = {1},
  pages     = {113--116},
  year      = {2002},
  doi = {10.1109/ICME.2002.1035731},
  publisher = {IEEE}
}

@inproceedings{martinez2012speaker,
  author    = {Martinez, Jorge and Perez, Hector and Escamilla, Enrique and Suzuki, Masahisa Mabo},
  title     = {Speaker recognition using {M}el frequency {C}epstral {C}oefficients ({MFCC}) and {V}ector quantization ({VQ}) techniques},
  booktitle = {CONIELECOMP 2012, 22nd International Conference on Electrical Communications and Computers},
  pages     = {248--251},
  year      = {2012},
  organization = {IEEE},
  doi = {10.1109/CONIELECOMP.2012.6189918}
}

@inproceedings{paulus2010state,
  author = {Paulus, Jouni and Müller, Meinard and Klapuri, Anssi},
  title     = {State of the Art Report: Audio-Based Music Structure Analysis},
  booktitle = {International Society for Music Information Retrieval Conference},
  pages     = {625--636},
  year      = {2010},
  address   = {Utrecht},
  doi = {10.5281/zenodo.1417289}
}
\end{document}